\documentclass[prb,twocolumn,showpacs,groupedaddress]{revtex4}
\usepackage{colordvi,epsfig,color,amsmath,amssymb}

\begin{document}

\def\be{\begin{equation}}
\def\ee{\end{equation}}
\def\bee{\begin{eqnarray}}
\def\eee{\end{eqnarray}}
\def\sech{\mbox{sech}}
\def\e{{\rm e}}
\def\d{{\rm d}}
\def\L{{\cal L}}
\def\U{{\cal U}}
\def\M{{\cal M}}
\def\T{{\cal T}}
\def\V{{\cal V}}
\def\R{{\cal R}}
\def\kb{k_{\rm B}}
\def\tw{t_{\rm w}}
\def\ts{t_{\rm s}}
\def\Tc{T_{\rm c}}
\def\gs{\gamma_{\rm s}}
\def\tm{tunneling model }
\def\TM{tunneling model }
\def\tilde{\widetilde}
\def\Deltac{\Delta_{0\rm c}}
\def\Deltamin{\Delta_{0\rm min}}
\def\Emin{E_{\rm min}}
\def\tauc{\tau_{\rm c}}
\def\tauac{\tau_{\rm AC}}
\def\tauw{\tau_{\rm w}}
\def\taumin{\tau_{\rm min}}
\def\taumax{\tau_{\rm max}}
\def\de{\delta\varepsilon / \varepsilon}
\def\pF{{\bf pF}}
\def\pFAC{{\bf pF}_{\rm AC}}
\def\halb{\mbox{$\frac{1}{2}$}}
\def\with{\quad\mbox{with}\quad}
\def\und{\quad\mbox{and}\quad}
\def\za{\sigma_z^{(1)}}
\def\zb{\sigma_z^{(2)}}
\def\ya{\sigma_y^{(1)}}
\def\yb{\sigma_y^{(2)}}
\def\xa{\sigma_x^{(1)}}
\def\xb{\sigma_x^{(2)}}
\def\spur#1{\mbox{Tr}\left\{ #1\right\}}
\def\erwart#1{\left\langle #1 \right\rangle}
\newcommand{\bbbone}{{\mathchoice {\rm 1\mskip -4mu l}{\rm 1\mskip -4mu l}{\rm 1\mskip -4.5mu l}{\rm 1\mskip -5mu l}}}

\title{Ultraslow quantum dynamics in a sub-Ohmic heat bath}

\author{P. Nalbach and M. Thorwart}
\affiliation{Freiburg Institute for Advanced Studies (FRIAS), School of Soft Matter Research, Albert-Ludwigs-Universit\"at Freiburg, Albertstr.\ 19, 79104 Freiburg}

\date{\today}

\begin{abstract}
We show that the low-frequency modes of a sub-Ohmic bosonic heat bath generate an effective dynamical asymmetry for an intrinsically symmetric quantum spin$-1/2$. An initially fully polarized spin first decays towards a quasiequilibrium determined by the dynamical asymmetry, thereby showing coherent damped oscillations on the (fast) time scale of the spin splitting. On top of this, the dynamical 
asymmetry itself decays on an ultraslow time scale and vanishes asymptotically since the global equilibrium phase is symmetric. 
We quantitatively study the nature of the initial fast decay to the quasiequilibrium and discuss the 
features of ultraslow dynamics of the quasiequilibrium itself.  
The dynamical asymmetry is more pronounced for smaller values of the sub-Ohmic exponent and for lower temperatures, which emphasizes the quantum many-body nature of the effect.
The symmetry breaking is related to the dynamic crossover between coherent and overdamped relaxation of the spin polarization and is not connected to the localization quantum phase transition. In addition to this delocalized phase, we identify a novel phase which is characterized by 
 damped coherent oscillations in the localized phase. This allows for a sketch of the zero-temperature phase diagram of the sub-Ohmic spin-boson model with four distinct phases.
\end{abstract}

\pacs{03.65.Yz, 64.70.Tg, 75.40.Gb, 05.30.Jp}



\maketitle

\section{Introduction}
\label{intro}

Open quantum dynamics \cite{SpiBoLe1987,Weiss99} describes relaxation and decoherence of quantum systems due to fluctuations induced by their environment. The noise characteristics determines the particular form
of the relaxation dynamics and is specified by the spectral distribution $J(\omega)$
of the environmental modes. An important class is electromagnetic noise which can generate
simple frequency-independent damping.
This is described by the well-known Ohmic environment $J(\omega)\sim \omega^s$ with exponent $s=1$.
Even the simplest model of a quantum two-level system
(TLS), the Ohmic spin-boson model, already shows nontrivial features, such as, e.g., a quantum phase transition of
localization due to strong damping \cite{SpiBoLe1987,Weiss99}.

Recently, a different class of fluctuations has moved into the focus of
attention~\cite{SpiBoKe1996,Mielke2002,Bulla2003,SpiBoAn2007,LeHur2007,SpiBoWi2009,SpiBoAl2009}, in which the
low-frequency components are more strongly pronounced compared to the Ohmic case. This case is
characterized by the sub-Ohmic distribution $0<s<1$, and the increased density of states of the low-frequency bath modes leads to peculiar and unexpected phenomena. The sub-Ohmic spin-boson model has been investigated by the
flow equation technique \cite{SpiBoKe1996,Mielke2002}, which allows to calculate spectral properties by equilibrium correlation functions. The results indicate a first order phase transition from delocalization to localization at some finite value of the system-bath coupling \cite{SpiBoKe1996}.
Wilson's numerical renormalization group approach has allowed \cite{Bulla2003} to identify a line of
{\em continuous\/} boundary quantum phase transitions of the sub-Ohmic
spin-boson model for all cases $0<s<1$. By the same method, weakly damped coherent oscillations on short time scales have been observed
 (but not further discussed) \cite{SpiBoAn2007} in the localized phase for $s=1/2$.  
 Quantum entanglement of the spin and the bath has shown to be enhanced at the 
second order quantum phase transition, which is indicated by a cusp in the entropy of entanglement \cite{LeHur2007}.
A continuous time cluster scheme based on Quantum Monte Carlo simulations has been used
\cite{SpiBoWi2009} to determine the critical exponent at the quantum phase transition and to
clarify a debate on the correctness of the quantum-to-classical mapping of the sub-Ohmic spin-boson model for $s<1/2$. Recently, a novel numerical approach has been put forward
\cite{SpiBoAl2009}, which uses a sparse polynomial representation of the basis of the
system-bath Hilbert space and which
allows to obtain numerical results based on exact diagonalization.
These results were all obtained by numerical methods which do not rely on a perturbative approximation.
On the other hand, these results also indicated \cite{SpiBoAn2007} that a perturbative approach
\cite{Chin2006,Lu2007,Wa2009,Gan2009} is by construction restricted to short times only, where it typically captures the renormalized coherent tunneling, but has problems to describe the approach to thermal equilibrium in the long-time limit, see also
Ref.\ \onlinecite{Wong2007} for a related discussion. Finally, we mention the recent work of 
Wang et al.~\cite{Wa2009} who discussed both the localization transition and the crossover between oscillatory and overdamped behavior. 
They find for all exponents $0\le s<1$ that with increasing coupling strength {\em first\/} the oscillatory behavior becomes overdamped and only then (for larger couplings) the transition to localization follows.

Sub-Ohmic fluctuations play a central role for the
quantum behavior of nanomechanical devices~\cite{Se2007}
due to the coupling of flexural modes to many surface defects of the
resonator. Moreover, an effective sub-Ohmic spin-boson model is studied as the archetype local quantum
critical system, showing a quantum critical transition (QCT)~\cite{SI1}.
The nature of the QCT in heavy Fermion metals is crucial to understand quantum phase 
transitions in general~\cite{SI2,SI2a}.
An important experimental puzzle is whether a non-linear scaling of the dynamic spin susceptibility
$\chi(\omega)\sim (\omega/T)^\nu$ with $\nu<1$ occurs~\cite{SI3} (here, $\omega$ is the frequency), which allows for the experimental
identification of the QCT in the spin density wave. Following a five-dimensional $\phi^4$ field theory, the corresponding quantum critical point 
in three dimensions can be identified by a Gaussian fixed point. 
A violation of the linear scaling $\nu=1$ would indicate an interacting fixed point with spin damping being present, while 
a nonlinear scaling  $\nu<1$ follows from assuming a non-interacting fixed point (with spin damping being absent). 
The question of the deviation from linear scaling is at present not finally answered \cite{SI3}.  
Moreover, the sub-Ohmic spin-boson model applies to
superconducting nanoscale devices~\cite{Sub1,Sub2}, such as qubits in contact with RLC transmission
lines~\cite{SpiBoTo2006}. Finally, the limit of $s \to 0$ corresponds to the
important class of $1/f-$noise \cite{Sub1,MussManNochSuchen,Paladino}. Often $1/f-$noise is generated by two level fluctuators as known from glasses~\cite{Phillips87} where they cause ultra slow (aging) dynamics\cite{TSGLNEQLu2003,TSGLNEQNa2004,TSGLNEQNa2005,TSGLNEQRo2003}. Due to the dominant low-frequency modes,
the dynamics is notoriously difficult to determine,
since strong non-Markovian effects even for relatively weak coupling arise.

In this work, we show that a sub-Ohmic heat bath generates an effective dynamical asymmetry for an
intrinsically symmetric quantum TLS (spin$-1/2$),
which leads to a {\it transient shifted quasiequilibrium} for the TLS, which then
itself decays extremely slowly.
This transient symmetry breaking occurs in the delocalized phase of the TLS, which
 follows the slowly changing quasiequilibrium. Moreover, it depends on the precise
form of the initial condition and only occurs for the case when the spin$-1/2$ is initially
shifted and the bath is initially thermalized to the shifted spin.
We establish a simple qualitative picture based on a perturbative analysis in terms of 
an asymmetric spin-boson system in order to interpret the dynamics of the spin polarization
at short times. There, the dynamical asymmetry shows an initial fast exponential decay 
towards a quasiequilibrium with a finite quasistatic asymmetry. 
Since the system is globally symmetric, a subsequent 
decay towards global thermal symmetric equilibrium follows 
in which the dynamical asymmetry vanishes as expected, as long as the system is in the 
delocalized phase. Our numerical data allow us to conclude that the 
time scale of this final decay is much larger than any simple 
time scale, such as given by the inverse spin splitting $\Delta^{-1}$, the inverse temperature $T^{-1}$ or the inverse cut-off frequency $\omega_c^{-1}$ of the bath. Moreover, we find that it is much larger than the observed decoherence time $\gamma_2^{-1}$. The  
asymptotic regime cannot be described anymore in terms of a simple perturbative picture. 

The feature of the ultraslow relaxation
dynamics together with the fast initial decay towards a quasiequilibrium resembles qualitatively the dynamical response of coarsening or a spinglass-like dynamics \cite{Leticia} to some extent. 
We should, however, emphasize that the numerical scheme is limited to time ranges which are too short  
to study the full asymptotic nature of the ultraslow decay, apart from stating its existence and 
some central features, such as its dependence on the model parameters. A full study of quantum critical features 
lies beyond the scope the approach. In general, the ultraslowly decaying transient asymmetry is shown to be more
pronounced for stronger system-bath coupling and smaller exponents $s$, which clearly 
illustrates the quantum many-body nature of the dynamical asymmetry.
In addition to this, we find {\em coherent oscillations\/} in the nonequilibrium polarization of the TLS in the
{\em localized\/} phase where in equilibrium all coherence is suppressed. These findings allow us
to sketch the zero-temperature phase diagram with four distinct phases. We argue that our results may have
immediate experimental implications, since the slow transient asymmetry will dominate all
spin fluctuations and thus the true equilibrium spin fluctuations, which are relevant for quantum critical behavior and which require asymptotically long times, may be difficult to access experimentally.
In turn, the slow dynamical asymmetry should also be taken into account by quantum engineering of nanodevices.

To structure the paper, we introduce the model in the next section and then discuss briefly the necessary technical modifications of the numerical exact quasi adiabatic propagator path-integral to tackle the spin dynamics with factorized and shifted or polarized bath initial conditions. 
In Section \ref{sec.decay}, we develop  our simple qualitative picture which quantitatively describes the formation of the 
dynamically generated asymmetry for the spin at intermediate times. In Sec.\ \ref{sec.phase},  
we sketch a complete phase diagram of the sub-Ohmic spin-boson model which summarizes our findings and which 
includes four distinct phases, before we finish with the conclusions.

\section{The model}

The spin-boson Hamiltonian is given by ($\hbar=1$), $H=H_S+H_{SB}+H_B$ with $H_S=\halb\Delta\sigma_x-\halb\epsilon\sigma_z$ with tunneling element $\Delta$ and
 asymmetry $\epsilon$. Moreover, $H_{SB}=-\frac{\sigma_z}{2}\sum_k\lambda_kq_k$ and
$H_B=\halb\sum_k \left(p_k^2+\omega_k^2q_k^2 \right)$,
where $q_k$ and $p_k$ are the position and momentum operators of the bath normal mode with wave vector $k$.
The bath is completely described by the spectral function
\be\label{spectrum} J(\omega) \,=\, \sum_k\frac{\lambda_k^2}{2\omega_k}\delta(\omega-\omega_k) \,=:\, 2\alpha\omega_c^{1-s}\omega^se^{-\omega/\omega_c}
\ee
with the bath cut-off frequency $\omega_c$ and the dimensionless coupling $\alpha$. In this work, we set $\omega_c=10 \Delta$~\cite{foot5}. The time dependent polarization $P(t):=\langle\sigma_z\rangle(t)$, after initial full polarization, $P(0)=1$, shows coherent damped oscillations and for an asymmetric two-level system additionally an exponential decay.
In the Ohmic case, $s=1$, the model displays a zero temperature quantum phase transition at $\alpha_c=1$ between a delocalized phase, where the spin can tunnel between its two states, and a localized phase.
Even in the delocalized phase at $\alpha<\alpha_c$, coherent dynamics is found only at
weak coupling $\alpha<\halb$. For $\halb\le\alpha<1$, overdamped relaxation occurs with a
dynamic crossover at $\alpha=\alpha_d=\halb$. In the sub-Ohmic regime, both a {\em continuous} (in $s$) localization transition and dynamic crossover are known~\cite{SpiBoKe1996,SpiBoAn2007,SpiBoWi2009,SpiBoAl2009,Wa2009}, but the interplay of both over the full range of exponents $s$ has not been discussed and its phase diagram is unidentified.

To determine $P(t)$
, we use the numerically exact real-time
quasiadiabatic propagator path integral (QUAPI)
\cite{QUAPI1,QUAPI2}. We are numerically
restricted to finite times and cannot investigate the localization
quantum phase transition, but rather the dynamic crossover between oscillatory and overdamped dynamics.
We calculate $P(t)$ for a symmetric TLS ($\epsilon=0$) in the delocalized phase
for three different exponents $s=0.25$, $0.5$ and $0.75$ and
focus mainly on low temperatures.
We study the coupling range from weak to strong (delocalized/localized phase)
as long as the dynamics is oscillatory.
\begin{figure}[t]
\epsfig{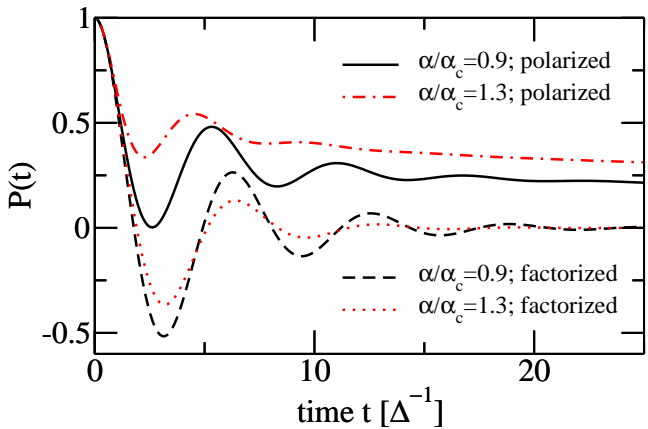}
\caption{\label{fig1} Polarization $P(t)$ for $T=0.1\Delta$, $s=0.25$ and $\omega_c=10\Delta$ for factorized and polarized bath initial conditions.}
\end{figure}
\section{QUAPI for the shifted bath initial condition}

The dynamics of the sub-Ohmic spin-boson model is
rather sensitive on the way how the system and bath are initially prepared, in particular with respect to their mutual interaction. It enters the formalism when the bath degrees of freedom are averaged over. To understand this, consider an experiment where a symmetric spin is in thermal equilibrium with the bath at temperature $T$ and at initial time $t=0$, the spin density operator $\rho_S$ is prepared such that $\langle\sigma_z\rangle(t=0) =1$. Then, the bath is in thermal equilibrium to
$\langle\sigma_z\rangle(t'<0) =0$, resulting in the factorizing initial condition for the total density operator at initial time, $\rho_0=\rho_S\otimes\rho_{B0}$ with $\rho_{B0} = \exp(-H_B/T)/Z_B$ with $k_B=1, H_B=\halb\sum_k \left(p_k^2+\omega_k^2q_k^2 \right)$ and $Z_B={\rm Tr } \exp(-H_B/T)$. A different case is to start preparing the spin long time before the experiment starts in order for the environment to thermalize to the {\em shifted\/} spin. This results in the shifted (correlated) initial condition $\rho_0=\rho_S\otimes\rho_{Bq}$ with $\rho_{Bq}= \exp(- \{H_B+H_{SB}|_{\sigma_z=1})/T\}/Z_{Bq}$, where $H_{SB}=-\frac{\sigma_z}{2}\sum_k\lambda_kq_k$ and
$Z_{Bq}= {\rm Tr } \exp(- \{H_B+H_{SB}|_{\sigma_z=1})/T\}$. All recent studies~\cite{SpiBoKe1996,SpiBoAn2007,SpiBoWi2009,SpiBoAl2009} were based on the shifted initial condition.

The coupling between the spin and a single bath mode scales with $1/\sqrt{N}$, with $N$ being the number of modes and thus is infinitely small for a continuum of modes. Therefore the difference between $\rho_{B0}$ and $\rho_{Bq}$ is small and believed to have little influence on the spin dynamics itself. This assumption is at the heart of open quantum dynamics since only an environment which is large enough to stay unaffected by a small quantum system can be averaged over without loss of crucial physics.
Lucke {\em et al.\/} \cite{SpiBoLu1997} showed that for an Ohmic bath, both initial conditions lead to sizeable differences for the spin dynamics only on very short times of the order of $1/\omega_c$. Thus, the spin dynamics which lives on a time scale of $1/E$, is unaltered as long as $E\ll\omega_c$.

The relevant parameter to distinguish both cases is the energy difference between the two initial conditions of the bath. It is estimated by the reorganization energy $E_R=\int_0^\infty d\omega \frac{J(\omega)}{\omega}$ and is dominated by the low frequency modes. As they are weakly present in the Ohmic case, their influence is
negligible when $\Delta\ll\omega_c$. In contrast, for a sub-Ohmic bath, any reorganization is {\it slow\/}, implying that the bath initial condition has a crucial influence on all time scales of the system dynamics.

In order to treat this situation properly, we adopt the QUAPI scheme, which has originally been formulated for factorizing initial conditions \cite{QUAPI1}, also to correlated initial conditions. QUAPI is an efficient, iterative and numerically exact method,
in which the real-time quantum evolution operator is sliced into $\cal N$ Trotter slices of
duration $\Delta t$. For the factorizing initial condition, the bath influence is captured by influence functional
\be
 I_0 = \exp\left\{ -\sum_{k=0}^{\cal N}\sum_{k'=0}^k \, [s_k^+-s^-_k]
\left[\eta_{kk'}s_{k'}^+-\eta_{kk'}^\star s^-_{k'}\right] \right\} \nonumber
\ee
with the bath correlators $\eta_{kk'}$ defined in Ref.~\onlinecite{QUAPI1} and the constant spin coordinate $s_k=\halb\sigma_z=\pm\halb$ at the time interval $[(k-\halb)\Delta t,(k+\halb)\Delta t]$. At any finite temperature, all bath correlations decay exponentially fast at asymptotically long times
\cite{QUAPI1}, which allows for a truncation of the bath-induced memory time beyond a certain time span. This corresponds in the above sum to setting $\eta_{kk'}=0$ when $|k-k'|>k_{max}$ for a given memory length $k_{max}$. Finally, convergence with respect to increasing $k_{max}$ has to be ensured. Inside the finite size memory time window, all correlations are taken into account exactly and the corresponding path sum is carried out deterministically.

For the shifted bath initial condition, we follow Ref.\ \onlinecite{SpiBoEg1994} to modify the influence functional $I_q=I_0 \exp(-\Phi_q)$ with
\[
\Phi_q = -iq\int_0^tdt'\,[s(t')-s'(t')] \int_0^\infty \hspace*{-2mm}d\omega \frac{J(\omega)F(T,\omega,t')}{\sinh(\omega/2T)}
\]
with $F(T,\omega,t')=\int_0^{1/T}d\tau \sinh [\omega/(2T)+i(t'+i\tau)]= 2\sinh[\omega/(2T)]\cos(\omega t')/\omega$, the spin $s(t)=\halb\sigma_z(t)$ and the shift $q=\halb$ for the shifted initial condition.
After Trotter slicing, we obtain with
$\eta_{q,k}(t) = \int_{(k-1)\Delta t}^{k\Delta t}
d\omega \frac{J(\omega)}{\omega^2} \sin(\omega t)$
\[ \Phi_q = -iq \sum_k[s_k^+-s_k^-] \eta_{q,k} \, .
\]
\begin{figure}[t]
\epsfig{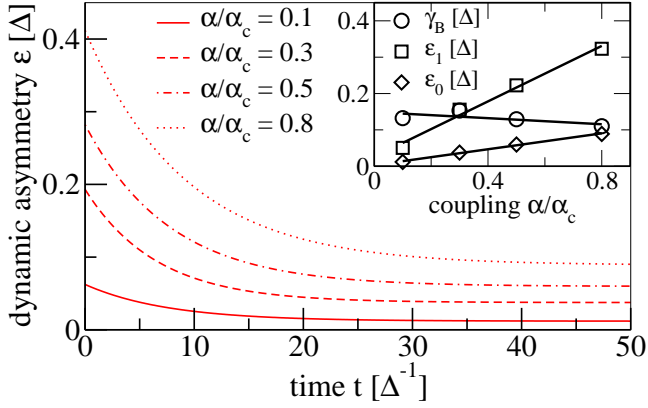}
\caption{\label{fig2} Dynamically generated asymmetry $\epsilon(t)$ for $T=0.1\Delta$, $s=0.5$ and $\omega_c=10\Delta$ for the polarized bath initial condition. The exponential decay is clearly visible by the semilogarithmic scale. Inset: Parameters extracted from a fit to Eq.\ (\ref{Fit}) as a function of $\alpha$ (symbols); solid lines are linear fits.}
\end{figure}

The bath correlators $\eta_{kk'}$ are all two-time correlation functions between the times $k'\Delta t$ and $k\Delta t$.
In contrast, $\eta_{q,k}$ describes a correlation between time $k\Delta t$ and the initial time $t=0$.
Within the QUAPI scheme, neglecting all correlators with $|k-k'|>k_{max}$ would imply to also neglect $\eta_{q,k}$
at times $t>k_{max}\Delta t$ and no sizeable difference for the two initial conditions would be found. However,
since $\eta_{q,k}$ depends only on a single running time index $k$, we can incorporate it completely by calculating the necessary $\eta_{q,k}$ in each time step during the iteration. Thus, the shifted bath correlations are taken fully into account, independent of $k_{max}$.

The two initial preparations induce qualitatively different behaviors of the dynamics of the sub-Ohmic spin-boson model. Fig.\ \ref{fig1} shows the polarization $P(t)$ for the factorizing initial condition for a symmetric spin ($\epsilon=0$) for different coupling strengths for $T=0.1\Delta$ and $s=0.25$ ($\alpha_c=0.022$~\cite{SpiBoWi2009}).
We find damped oscillations where damping grows with increasing $\alpha$. As expected from a Bloch-like picture, the oscillations occur around zero average. Surprisingly even in the localized phase, $P(t)$
shows quantum coherent oscillations out of equilibrium. 
In the localized phase the bath modes are too slow to localize the out-of-equilibrium TLS fast enough and thus damped coherent oscillations still occur at short times.

In contrast to that, Fig.\ref{fig1} also shows $P(t)$ for the shifted bath initial condition. As before, we find damped oscillations with a damping constant increasing with $\alpha$. However, in striking qualitative difference, the oscillations occur around a finite value $P_{\rm av}>0$, even for coupling strengths well in the delocalized phase, which indicates an intermittent symmetry breaking. Note that neither $P_{\rm av}>0$ nor the long-time value $P(t=25/\Delta)$ nor the occurrence of oscillations allow to distinguish between the delocalized and the localized phase. The latter results are in line with those of Ref.\ \onlinecite{SpiBoAn2007}. The dependence on the initial preparation, however, has not been disclosed before.

\section{Ultraslow decay of the effective asymmetry}
\label{sec.decay}
\begin{figure}[t]
\epsfig{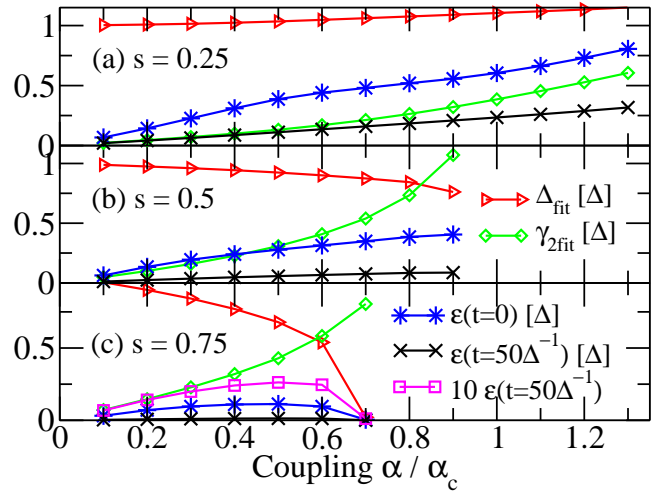}
\caption{\label{fig3} Initial ($\epsilon(t=0)=\epsilon_0$)
and final ($\epsilon(t=50/\Delta)$) dynamical asymmetry, tunneling matrix element
$\Delta_{\rm fit}$ and decoherence rate $\gamma_{\rm 2, fit}$ at $T=0.1\Delta$ for the polarized bath initial condition for $s=0.25$ (a), $0.5$ (b) and $0.75$ (c). Note that in (c) the data (magenta squares)
for $\epsilon(t=50/\Delta)$ have been scaled up to increase visibility.}
\end{figure}

In order to study the effect of the shifted oscillations more quantitatively, they can be formalized in terms of an effective time-dependent asymmetry $\epsilon(t)$ for the TLS which generates a finite polarization $P_{\rm av}$.
The polarization of a TLS with a static asymmetry $\epsilon$ at low temperatures and weak coupling is given~\cite{Weiss99} by
\bee\label{Fit} P(t) &=& \left(\frac{\Delta}{E}\right)^2 \{ \cos(Et) +\frac{\gamma_2}{E} \sin(E t) \} e^{-\gamma_2t} \nonumber \\
&&+ \left(\frac{\epsilon}{E}\right)^2 e^{-\gamma_{\rm 1} t} -\left(\frac{\epsilon}{E}\right) (1-e^{-\gamma_{\rm 1} t}) \, ,
\eee
with $E=\sqrt{\Delta^2+\epsilon^2}$. Certainly, for our intrinsically symmetric case, 
at asymptotically long times, the asymmetry has to vanish, while our data reveal that the asymmetry is generated 
at short to intermediate times, rendering $\epsilon$ effectively time-dependent. Nevertheless, we expect that particularly at short times, the dynamics is well described by a lowest order perturbative description in the tunneling strength, yielding an initial exponential decay also for the bath correlations. 
Thus, it is suggestive to use the heuristic approach $\epsilon \to \epsilon(t)=\epsilon_0+\epsilon_1\exp(-\gamma_B t)$ for the dynamical asymmetry and $E=E(\epsilon(t))$ in Eq.\ (\ref{Fit}). Clearly, this approach breaks down at asymptotically long times since it would yield 
a finite static asymptotic bias $\epsilon(t\to \infty) = \epsilon_0$, which cannot occur in a globally symmetric system. 
Nevertheless, we use this heuristic approach to show the features of the dynamical asymmetry on a qualitative level. 

Fitting our data for $P(t)$ at short to intermediate times with Eq.\ (\ref{Fit}) allows us to extract $\epsilon_0, \epsilon_1$ and $\gamma_B$. Fig.~\ref{fig2} shows the result of the fit for $\epsilon(t)$ for $T=0.1\Delta$, for different $\alpha$ and $s=0.5$ with critical coupling for localization $\alpha_c\simeq 0.1065$~\cite{SpiBoAn2007}. The results of the fits with the effective dynamical asymmetry 
are perfect (not shown) on the considered time span up to $t=50$. This approach will no longer hold for truly asymptotic times 
$t\to \infty$. With increasing $\alpha$, the dynamical asymmetry (both at $t=0$ and $t=50\Delta^{-1}$) increases, but its short time decay rate $\gamma_B$ decreases slightly. The inset of Fig.~\ref{fig2} shows the extracted parameters vs.\ $\alpha$ and linear fits to the data.

A simple qualitative picture can be obtained by assuming that the initially shifted bath modes generate this transient dynamical asymmetry. The slow modes drag the TLS to its original position, similar to the cage effect in the fractional Langevin equation~\cite{Sub3}. This causes a non-zero effective polarization $P_{\rm av}\ne 0$ and a finite asymmetry both of which are clearly dynamically generated by the sub-Ohmic bath correlations. In general,
the symmetry breaking is enhanced with increasing coupling, emphasizing the many-body character of the asymmetry.
At asymptotically long times, we expect the bath to relax slowly towards global equilibrium with the TLS being in either of its two symmetric eigenstates in which $\langle\sigma_z\rangle_{eq}=0$ and the asymmetry will thus slowly vanish. The decay of the initial 
coherent oscillations shown in Fig.~\ref{fig1} occurs on a time scale of the same order as the period of the oscillation, but is 
very fast compared to the decay of $P_{\rm av}$ which we actually cannot fully observe but clearly conjecture~\cite{foot2}. 
This effect depends on the initial preparation of the bath and, moreover, it is absent in an Ohmic bath \cite{SpiBoLu1997}. We note that the feature of an initial decay with a time scale of the order of a {\it microscopic} decay rate of the spin and a subsequent ultraslow decay towards a global equilibrium resembles that of coarsening or of spin glasses \cite{Leticia} after a quench~\cite{foot1}. Although the sub-Ohmic model has no explicit spatial disorder, its low-frequency noise implicitly mimics disorder, similar to $1/f$-noise. The latter results from localized fluctuators with broadly randomly distributed parameters~\cite{MussManNochSuchen}.
The slow transients might render the sub-Ohmic spin-boson model a toy model to study quantum glasses~\cite{SpiBoCu2002}.
\begin{figure}[t]
\epsfig{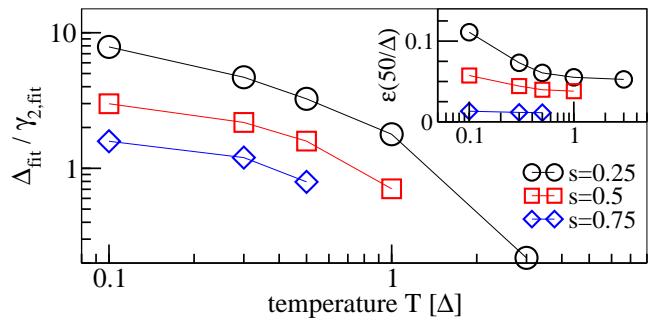}
\caption{\label{fig4} Main: Ratio of tunneling element $\Delta_{fit}$ and decoherence rate $\gamma_{2,fit}$ vs. $T$ for
 $\alpha/\alpha_c=0.5$.
Inset: The corresponding final asymmetry
$\epsilon(t=50/\Delta)=\epsilon_{0,fit}+\epsilon_{1,fit}\exp(-\gamma_{B,fit} t)$ vs.\ temperature.}
\end{figure}
%


Fig.~\ref{fig3} presents results for the fit of 
the initial dynamical asymmetry $\epsilon(t=0)=\epsilon_0+\epsilon_1$, the asymmetry $\epsilon(t=50/\Delta)$ for the maximal 
simulation time, and the tunneling element $\Delta_{\rm fit}$ and decoherence rate $\gamma_{2,\rm fit}$ vs.\ $\alpha/\alpha_c$ for $s=0.25, 0.5$ and $s=0.75$, with
$\alpha_c\simeq0.022, 0.1065$ and $\alpha_c\simeq 0.3$~\cite{SpiBoWi2009}, respectively. Note that we have set $\gamma_{1}=2\gamma_{2}$, which is strictly valid for a symmetric system \cite{Weiss99} and that we have ignored the additional correction term for the rate due to a finite asymmetry. All data show the initial oscillatory behavior except for $s=0.75$ and $\alpha= 0.7\alpha_c$ which exhibits overdamped dynamics.
For $s=0.25$, the asymmetry is smoothly increasing for larger $\alpha$ up to the phase transition and even beyond. At the same time, the tunneling matrix element is hardly influenced and larger than the decoherence rate which increases with coupling. Thus, the presence (absence) of oscillations is not a sufficient condition for the existence of the delocalized (localized) phase.
For $s=0.5$, the asymmetry is generally smaller, but also increases with increasing $\alpha$. For $\alpha\ge 0.9\alpha_c$, the decoherence rate exceeds the tunneling element hinting at the crossover to overdamped behavior for stronger couplings.
For $s=0.75$, the dynamical asymmetry is small, depends nonmonotonously on $\alpha$ and vanishes for $\alpha\ge 0.7\alpha_c$. The tunneling matrix element also vanishes there, indicating the crossover to overdamped dynamics. Thus both the overdamped dynamics and the dynamical asymmetry are closely connected.

Thus, we can summarize our physical picture in the following way: 
In general, when the TLS is fully polarized and released at $t=0$, it starts to oscillate coherently and resonant bath modes damp the oscillations. Since the slow bath modes are not yet in equilibrium with the evolving TLS, but still shifted, they cause a drag for the TLS to its initial state, resulting in the formation of a dynamical asymmetry. This decays to zero on an ultraslow time scale since the global equilibrium is symmetric. Thus, the local quantities decay first to a {\it local\/} quasiequilibrium generated by the immediate environment. This oscillatory decay at short to intermediate times is well characterized by exponential decay rates $\gamma_{1/2}$ (see Eq.\ (\ref{Fit}), data not shown explicitly) and a 
dynamical asymmetry $\epsilon(t)$, which itself varies with time. Following our simple perturbative line of reasoning,   $\epsilon(t)=\epsilon_0+\epsilon_1\exp(-\gamma_B t)$ consists of two parts. One part, $\epsilon_1\exp(-\gamma_B t)$, decays exponentially 
with rate $\gamma_B$ and quickly vanishes as time increases. The second part $\epsilon_0$ decreases only very weakly with growing time which is revealed by our numerical data, but has to vanish in the asymptotic limit $t \to \infty$ due to global symmetry. 
Hence, only at very large times, a global equilibrium state emerges. At stronger coupling, strong correlations between the TLS and resonant modes generate an overdamped dynamics. The responsible modes are not the slow modes, thus the dragging effect is suppressed.

\section{Dynamic cross-over and sketch of a phase diagram}
\label{sec.phase}

With increasing temperature, the dynamics becomes overdamped as well. Fig.~\ref{fig4} shows the ratio $\Delta_{\rm fit}/\gamma_{\rm 2,fit}$ for $\alpha/\alpha_c=0.5$. In general, it decreases with increasing $T$. Only for $s=0.25$ at $T=3\Delta$, we observe overdamped behavior. The data shown in Fig.~\ref{fig4} indicate that the crossover temperature at which the overdamped regime start increases with decreasing exponent $s$ (at fixed relative coupling $\alpha/\alpha_c=0.5$). The final asymmetry (inset of Fig.~\ref{fig4}) decreases with temperature in a monotonic way, also around the crossover temperature.
\begin{figure}[t]
\epsfig{file=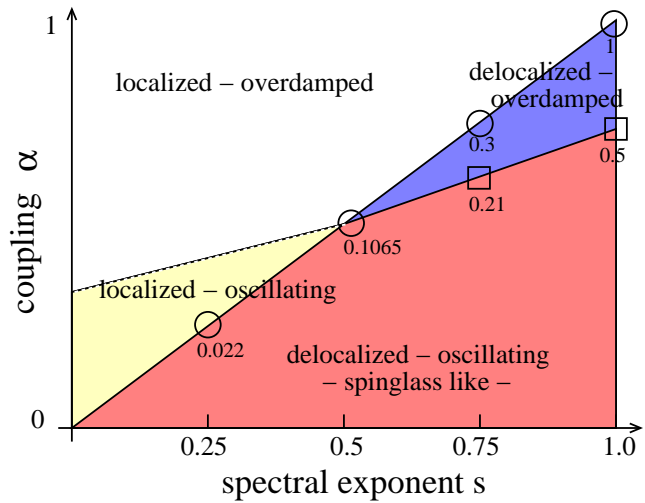,width=8.5cm}
\caption{\label{fig5} Sketch of a phase diagram of the sub-Ohmic spin-boson model at $T=0$, see text.}
\end{figure}

The findings can be summarized in a phase diagram
sketched in Fig.\ref{fig5} (at $T=0$). The black circles mark the equilibrium localization
phase transition known in the literature~\cite{SpiBoWi2009} (note that the numbers refer
to the values of $\alpha$ along the ordinate, the scale, however, is not linear).
The second solid line with the squares mark the dynamic crossover
from coherent to overdamped dynamics~\cite{Lu2007,foot6}.
The delocalized (weak-coupling or symmetric) phase is reflected in the coherent
oscillations of $P(t)$, in which we have unexpectedly found a transient symmetry breaking for all $s<1$.
For $1/2 \le s\le1$, a dynamical crossover to the overdamped delocalized phase occurs without
transient symmetry breaking, before the localized phase is reached for large $\alpha$.
Unexpectedly, for $s\le 1/2$, no overdamped dynamics occurs but the nonequilibrium
polarization still shows quantum coherent (damped) oscillations in the localized phase, where in equilibrium
coherence is completely suppressed. Although we could not obtain converged results for stronger couplings, we expect
that the oscillatory dynamics deeper in the localized phase should vanish.

The phase diagram summarizes our two major results: (i) In the delocalized phase where the dynamics is oscillatory,  the initial preparation of the bath strongly influences the dynamics. Thus, the dynamics in this phase is dominated by the ultra-slow sub-Ohmic bath modes. Any experiment addressing this phase will have to deal with these unusual long-time transients and true equilibrium quantities will be hard to access. Accordingly, one cannot expect to observe the spectrum of the equilibrium fluctuations when approaching the localization phase transition from this phase. This has to be taken into account when analyzing the data for the dynamic structure factor of heavy fermion systems within a picture of local quantum criticality. The latter assumes the local critical fluctuations to be of sub-Ohmic nature. (ii) Moreover, the phase denoted as ``localized--oscillatory'' in Fig.\ \ref{fig5} is characterized by an oscillating polarization, although the spin is in the localized phase where in equilibrium any oscillation is suppressed. This phase is novel and, to the best of our knowledge, has not been revealed or discussed before. Note that it is not an equilibrium phase. Strictly speaking, it is just a part of the equilibrium phase denoted as ``localized'', but from the viewpoint of the real-time dynamics, the two regions (one with oscillatory dynamics and one without) are distinct. It is known that for $s>1/2$ no oscillatory dynamics occurs for couplings stronger than the critical coupling of the 
crossover to overdamped dynamics. Accordingly, in the localized phase (which occurs here at even stronger couplings),  no oscillatory dynamics is observed. We believe that even for $s<1/2$, the oscillatory behavior will be suppressed at very large couplings. Hence, we conjecture the dashed phase separation line between localized-oscillatory and the localized-overdamped region in Fig.\ \ref{fig5}. 

We note that experimental studies in this phase might observe a dynamical behavior which does not allow to distinguish it from behavior found in the delocalized phase unless the experimental time scale exceed the time scale of the ultraslow decay in the delocalized phase since in the localized phase the asymmetry will not decay at any time. Although the numerical scheme does not allow to observe the ultraslow decay, we can estimate it to be at least two orders of magnitude slower than the intrinsic time scale $\Delta^{-1}$ of the spin as we observed finite asymmetries up to times $50 \Delta^{-1}$ (see Fig.~\ref{fig2}).

\section{Summary}

To summarize, we have found a distinct ultraslow transient dynamics of the polarization
in the sub-Ohmic spin-boson model. A dynamically correlation-induced
asymmetry is generated, which mimics a local quasiequilibrium and generates a symmetry broken phase at intermediate times. It occurs due to the dominating low-frequency bath modes and only for an initially polarized thermal bath.
The latter refers to the case when the system is initially prepared in a non-thermal polarized state and the bath
can thermalize to that before relaxation to the global thermal equilibrium can start. Technically, this kind of initial preparation can be included by adopting the numerical quasiadiabatic propagator path integral scheme. For this, an additional term in the Feynman-Vernon influence phase is included which only depends on the initial and the momentary observation time, but not on the times in between. Hence, the general strategy of a finite memory time in QUAPI is not affected and this term can be taken into account in an exact manner.
With that, we can show that the dynamic crossover between oscillatory and
overdamped behavior is connected to the transient symmetry breaking and
not to the localization quantum phase transition.

Finally, we comment on an important possible consequence of our findings for identifying the quantum
critical transition in heavy Fermion systems~\cite{SI2,SI2a}. An effective sub-Ohmic spin-boson model is used for
this~\cite{SI1}. At present , the non-linear scaling of the dynamic susceptibility
$\chi(\omega)\sim (\omega/T)^\nu$ with $\nu<1$ is used as a criterion~\cite{SI3}. However, our results
suggest that the ultraslow relaxation dynamics in the sub-Ohmic
spin-boson model requires to carefully measure the relaxation dynamics at very long times. In fact,
this could render an experimental observation of
the equilibrium localization quantum phase transition difficult in general. For the same reason,
the nonlinear scaling of the dynamical susceptibility has to be evaluated very carefully in order
to ensure the existence of local quantum criticality in heavy Fermion systems.

\acknowledgments

We thank K. Binder, H.\ Grabert and S. Kehrein for discussions and
acknowledge support by the Excellence Initiative of the German Federal and State Governments.


\begin{thebibliography}{99}

\bibitem{SpiBoLe1987} A.J. Leggett, S. Chakravarty, A.T. Dorsey, M.P.A. Fisher, A. Garg, W. Zwerger,
Rev. Mod. Phys. {\bf 59}, 1 (1987).

\bibitem{Weiss99} U. Weiss, {\it Quantum Dissipative Systems} (3rd ed., World Scientific, Singapore, 2007).

\bibitem{SpiBoKe1996} S. Kehrein and A. Mielke, Phys. Lett. A {\bf 219}, 313 (1996).

\bibitem{Mielke2002} T. Stauber and A. Mielke, Phys. Lett. A {\bf 305}, 275 (2002).

\bibitem{Bulla2003}R. Bulla, N.-H. Tong, and M. Vojta, Phys. Rev. Lett. {\bf 91}, 170601 (2003).

\bibitem{SpiBoAn2007} F.B. Anders, R. Bulla, and M. Vojta, Phys. Rev. Lett. {\bf 98}, 210402 (2007).

\bibitem{LeHur2007} K. Le Hur, P. Doucet-Beaupre, and W. Hofstetter, Phys. Rev. Lett.
 {\bf 99}, 126801 (2007).

\bibitem{SpiBoWi2009} A. Winter, H. Rieger, M. Vojta, and R. Bulla, Phys. Rev. Lett. {\bf 102}, 030601 (2009).

\bibitem{SpiBoAl2009} A. Alvermann and H. Fehske, Phys. Rev. Lett. {\bf 102}, 150601 (2009).

\bibitem{Chin2006} A. Chin and M. Turlakov, Phys. Rev. B {\bf 73}, 075311 (2006).

\bibitem{Lu2007} Z. L\"u and H. Zheng, Phys. Rev. B {\bf 75}, 054302 (2007).

\bibitem{Wa2009} Q. Wang, A.-Y. Hu, and H. Zheng, Phys. Rev. B {\bf 80}, 214301 (2009).

\bibitem{Gan2009} C. Gan and H. Zheng, Phys. Rev. E {\bf 80}, 041106 (2009).

\bibitem{Wong2007} H. Wong and Z.-D. Chen, Phys. Rev. B {\bf 76}, 077301 (2007).

\bibitem{Se2007} C. Seoanez, F. Guinea, and A.H. Castro Neto, Europhys. Lett. {\bf 78}, 60002 (2007).

\bibitem{SI1} S. Kirchner and Q. Si, Phys. Rev. Lett. {\bf 100}, 026403 (2008).


\bibitem{SI2} Q. Si, S. Rabello, K. Ingersent, and J.L. Smith, Nature {\bf 413}, 804 (2001).

\bibitem{SI2a} P. Gegenwart, T. Westerkamp, C. Krellner, Y. Tokiwa, S. Paschen, 
C. Geibel, F. Steglich, E. Abrahams, and Q. Si, Science {\bf 315}, 969 (2007).

\bibitem{SI3} P. Gegenwart, Q. Si, and F. Steglich, Nature Phys. {\bf 4}, 186 (2008).

\bibitem{Sub1} A. Shnirman, Y. Makhlin, and G. Sch\"on, Phys. Scr. {\bf T102}, 147 (2002).

\bibitem{Sub2} Y. Nakamura, Y.A. Pashkin, T. Yamamoto, and J.S. Tsai, Phys. Rev. Lett. {\bf 88}, 047901 (2002).

\bibitem{SpiBoTo2006} N.-H. Tong and M. Vojta, Phys. Rev. Lett. {\bf 97}, 016802 (2006).

\bibitem{MussManNochSuchen} J. Bergli, Y.M. Galperin, and B.L. Altshuler, New J. Phys. {\bf 11}, 025002 (2009).

\bibitem{Paladino}E. Paladino, L. Faoro, G. Falci, and Rosario Fazio, Phys.\ Rev.\ Lett.\ {\bf 88}, 
228304 (2002); E. Paladino, A. D'Arrigo, A. Mastellone, and G. Falci, Phys. Scr. {\bf T137}, 014017 (2009).

\bibitem{Phillips87} W.A. Phillips, Rep. Prog. Phys. {\bf 50}, 1657 (1987).

\bibitem{TSGLNEQRo2003} D. Rosenberg, P. Nalbach, and D.D. Osheroff, Phys. Rev. Lett. {\bf 90}, 195501 (2003).
\bibitem{TSGLNEQLu2003} S. Ludwig, D. Rosenberg, P. Nalbach, D.D. Osheroff, Phys. Rev. Lett. {\bf 90}, 105501 (2003).
\bibitem{TSGLNEQNa2004} P. Nalbach, D.D. Osheroff, S. Ludwig, J. of Low Temp. Phys. {\bf 137}, 395 (2004).
\bibitem{TSGLNEQNa2005} P. Nalbach, Phys. Rev. B {\bf 71}, 052201 (2005).

\bibitem{Leticia} L.F. Cugliandolo, Lecture Notes on {\it Dynamics of glassy systems}, arXiv:cond-mat/0210312v2.

\bibitem{foot5} Keep in mind that the critical coupling $\alpha_c\equiv\alpha_c(\Delta/\omega_c)$.

\bibitem{QUAPI1} N. Makri, D. E. Makarov, J. Chem. Phys. {\bf 102}, 4600 (1995); ibid.\ {\bf 102}, 4611 (1995);
 N. Makri, J. Math. Phys. {\bf 36}, 2430 (1995).

\bibitem{QUAPI2} M. Thorwart, P. Reimann, P. Jung, and R. F. Fox,
Chem. Phys. {\bf 235}, 61 (1998); M. Thorwart, P. Reimann, and P. H\"anggi, Phys. Rev. E {\bf 62}, 5808 (2000).

\bibitem{SpiBoLu1997} A. Lucke, C.H. Mak, R. Egger, J. Ankerhold, J. Stockburger, and H. Grabert, J. Chem. Phys. {\bf 107}, 8397 (1997).

\bibitem{SpiBoEg1994} R. Egger and C.H. Mak, Phys. Rev. B {\bf 50}, 15210 (1994).


\bibitem{Sub3} S. Burov and E. Barkai, Phys. Rev. Lett. {\bf 100}, 070601 (2008).

\bibitem{foot2} The only alternative to the conjectured ultraslow decay of the dynamical asymmetry towards zero is assuming that the system is already in the localized phase. Then, our results would show strikingly different critical coupling strengths $\alpha_c$ in comparison to other results~\cite{SpiBoWi2009,Bulla2003,SpiBoAn2007} obtained with different approaches. Therefore, the conjectured ultraslow decay of the dynamical asymmetry towards zero is the only sensible assumption.

\bibitem{foot1} $P(t)=P(t;t_p)$ is in fact a two-time correlator, involving the preparation time $t_p$. For the factorized initial condition, $t_p=0^-$, and for the polarized one, $t_p=-\infty$.

\bibitem{SpiBoCu2002} L.F. Cugliandolo, D.R. Grempel, G. Lozano, H. Lozza,
and C.A. da Silva Santos, Phys. Rev. B {\bf 66}, 014444 (2002).

\bibitem{foot6} The dynamic crossover occurs precisely at $\alpha=\alpha_d$, but there is no equilibrium phase transition.

\end{thebibliography}
\end{document}